\documentclass[conference]{IEEEtran}
\usepackage[T1]{fontenc}
\usepackage[utf8]{inputenc}
\usepackage[english]{babel}
\usepackage{cite}
\usepackage{amsmath,amssymb,amsthm}
\usepackage{graphicx}
\usepackage{booktabs}
\usepackage{array}
\usepackage{tabularx}
\usepackage{multirow}
\usepackage{url}
\usepackage{xcolor}
\usepackage{csquotes}
\usepackage{listings}
\usepackage{hyperref}

\title{Hagenberg Risk Management Process (Part 3):\\
Operationalization, Probabilities, and\\Causal Analysis}

\author{\IEEEauthorblockN{Eckehard Hermann, Harald Lampesberger}
\IEEEauthorblockA{University of Applied Sciences Upper Austria, Campus Hagenberg, Austria\\
Email: \{eckehard.hermann, harald.lampesberger\}@fh-hagenberg.at}}

\begin{document}
\selectlanguage{english}
\maketitle

\begin{abstract}
For risks that cannot be accepted, sufficiently mitigated, or eliminated, continuous observation to detect early signals is a viable approach but requires a model that can be operationalized.
The Hagenberg Risk Management Process bridges this gap between qualitative risk analysis, using contextualized polar heatmaps (triage), and realtime risk management by extending Bowtie diagrams into a formal probabilistic runtime model. We introduce Realtime Risk Studio, a domain‑specific modeling tool that (i) transforms Bowtie structures (causes, top event, barriers, consequences) into a directed acyclic graph (DAG) suitable for Bayesian inference, (ii) adds explicit safe‑state semantics, and (iii) designates Activation Nodes as intervention points.
Bowtie models are qualitative; however, Bayesian inference requires actual probabilities. As a second contribution, we present Probability Capture, a tool that complements our Realtime Risk Studio by automatically generating questionnaires from a DAG model so experts can provide estimates. The tool analyzes disagreement and aggregates conditional‑probability assessments using both descriptive dispersion analysis and prior‑regularized methods.
Causal analysis can then provide insights into the DAG model, for example, via d‑separation, adjustment‑set inspection, do‑calculus for what‑if analysis, local independence checks, evidence updating, and impact‑oriented searches for effective interventions.

This workflow is illustrated with an instant‑payments gateway scenario, demonstrating (a) explicit safe‑state semantics, (b) Bowtie‑to‑DAG operationalization, (c) probability capture with visible expert noise, and (d) causal what‑if analysis on a transformed and enriched model. Rather than presenting a statistical validation, the paper contributes a method and prototype system that transforms partially mitigated risks into observable, probabilistic, and intervention‑ready models for realtime risk management.
\end{abstract}


\section{Introduction}
The first two contributions of the Hagenberg Risk Management Process establish a layered perspective on the risk management of complex infrastructures. Part~1 extends classic heatmaps by explicit context dimensions and thus creates a formal basis for context-sensitive triage~\cite{hermann2026part1}. Part~2 connects this triage with a Bowtie-based case analysis, a structured capture of barriers, and a derivation into a directed acyclic graph (DAG) as the basis of a Bayesian network (BN)~\cite{hermann2026part2}. The present paper addresses the next gap: how can a semantically well-structured risk model be turned into a model that is observable in operation, probabilistically evaluable, and useful for interventions?

This question arises in particular when a risk is not organizationally accepted, yet at the same time cannot be fully eliminated or sufficiently mitigated. In such cases, documenting the risk once or discussing it qualitatively is not enough. Rather, the situation in question must be continuously observed during operations in order to detect early signals, integrate evidence, and react quickly in the event of occurrence or escalation. Regulatory and normative contexts such as DORA~\cite{dora2022} and continuous-monitoring concepts in the NIST environment~\cite{nist800137} underline precisely this need for ongoing situational awareness and control.

A classic Bowtie diagram is only partially suitable for this step. Its strength lies in the transparent, communicative structuring of threats, top event, preventive and mitigative barriers, and consequences~\cite{deruijter2016,iec31010}. This structure is valuable for workshops, governance, and audits. For realtime monitoring, however, a Bowtie lacks a decisive component: the explicit normal state. By definition, a Bowtie begins at the top event and typically considers consequences as paths \emph{after} the occurrence of the top event. In operation, however, most of the time the case \emph{Top Event = false} applies. Monitoring must therefore probabilistically model not only adverse paths, but especially the continued normality of the system and distinguish it from alternative damage states. Without this explicit representation, it remains unclear what evidence during normal operation actually refers to.

This observation leads to the core thesis of this paper: for operational observation, evidence processing, and decision support, a Bowtie must be transformed into a probabilistic model with an explicit normal state and a clean causal structure. A DAG and the Bayesian network built on top of it are suitable for this purpose because they bring together states, dependencies, evidence, and interventions in a common formal framework~\cite{koller2009,fenton2012,pearl2009}.

The paper therefore pursues four goals. First, the problem of the realtime use of Bowtie models is systematically elaborated and situated in the literature. Second, related work on DAGs, Bayesian networks, Pearlian causality, expert elicitation, and relevant tool ecosystems is summarized. Third, our tool contributions, Realtime Risk Studio and Probability Capture, are presented. Together they provide Bowtie-to-DAG transformation, expert questionnaires for quantitative probability estimates, and causal analysis as an integrated toolchain. Fourth, this chain is demonstrated on a domain-realistic example from the DORA-related instant-payments environment that we introduced in our previous paper~\cite{hermann2026part2}. The main scientific contribution lies in the integration of four elements into one operational workflow: explicit safe-state semantics on the consequence side, Bowtie-to-DAG operationalization with activation nodes as potential intervention points in causal analysis, structure-derived probability estimation with visible noise, and causal what-if analysis on the same runtime model. The contribution is therefore positioned as a {method and system paper}: the paper introduces a modeling pattern and a reference implementation, while the case study serves as an illustrative walkthrough of the workflow rather than as an empirical benchmark study.

\section{Bowtie Diagrams are Insufficient for Realtime Monitoring}
\subsection{Weak Runtime Basis}
The Bowtie method is established in high-risk and safety domains because it brings together causes, threats, top event, barriers, and consequences in an easily communicable representation~\cite{iec31010,deruijter2016}. The method is therefore particularly effective when risks are to be jointly analyzed in workshops, barriers discussed, and responsibilities made visible. In Part~2 of the Hagenberg Risk Management Process, exactly this strength is used: after context-sensitive triage, materialized risks are deepened in a Bowtie in order to generate an operational case analysis from a prioritized risk description~\cite{hermann2026part2}.

At the same time, the literature describes limits of the method. De Ruijter and Guldenmund~\cite{deruijter2016} point out that Bowtie is used inconsistently in many methodological variants and, in its simple form, is more of a structuring and communication aid than a strictly quantitative method. Khakzad et al.~\cite{khakzad2013} additionally show that classic Bowties have difficulties with conditional dependencies, probabilistic updates, and dynamic state change. This is particularly critical for realtime applications, because monitoring depends on two properties: observations must be incorporable into the model, and new evidence must update the risk assessment over time.

In addition, there is a semantic asymmetry in a Bowtie. The left-hand side models causes or preventive barriers prior to the top event; the right-hand side models consequences and mitigative barriers after the top event. This representation is useful for analysis and communication because it clearly separates preventive and mitigative measures. For runtime monitoring, however, it is insufficient because it does not treat the dominant operating case, \emph{the non-occurrence of the top event}, as an explicit outcome state. In operational monitoring, the relevant question is not only how damage unfolds after occurrence, but also whether the system remains in safe normal operation, which observations make this state plausible, and which evidence moves away from it.

\subsection{The Need for a Normal State}
At its core, a monitoring system does not merely answer the question \emph{``How likely is a loss?''}, but also \emph{``Which system state is currently present?''}. Without explicit modeling of the normal state, this question is formally incomplete. In a Bayesian network, each node must have a well-defined set of states whose probabilities sum to~1~\cite{koller2009,fenton2012}. For the result or consequence node, this means that if several adverse end states are modeled, an explicit normal state is needed to represent regular operation, that is, the case in which no loss-relevant path has occurred. In the present approach, this role is assumed by the state \emph{safe}.

This addition is an operational necessity to distinguish normal operation from adverse consequences. It creates a basis for interpreting evidence from monitoring sources in a meaningful way: stable telemetry, normal response times, and the absence of error signals support the \emph{safe state}; contrary evidence shifts the posteriors toward degraded or harmful end states. Explicit modeling of the \emph{safe state} allows non-accepted risks to be tracked over longer periods without artificially focusing on the top event all the time. The operational comparison is then \emph{safe versus adverse states} instead of \emph{consequence A versus consequence B}.

Several works from the Bowtie-BN community support the broader need for dynamic probabilistic risk assessment. Khakzad et al.~\cite{khakzad2012,khakzad2013} transform Bowties into Bayesian networks in order to integrate observations and dynamically adjust probabilities. Zurheide et al.~\cite{zurheide2021} operationalize Bowtie-to-BN transformation ideas in \emph{pyBNBowtie} and thereby show that Bowtie structures can indeed be transferred into a probabilistic modeling basis. The present argument goes one step further: for the monitoring semantics pursued here, the explicit modeling of the normal state is treated as a design requirement of the runtime model rather than as a claim that the earlier Bowtie software literature had already formalized.

\subsection{The Problematic Transition to Operations}
Many risk models do not fail because of the quality of the initial recording, but at the transition into operations. A Bowtie can support excellent workshops, management reviews, and audit trails. For operational use, however, four things are typically missing.

First, there is no robust semantics for evidence integration. Observations from logs, sensors, operational metrics, or alarms cannot be propagated natively in a Bowtie. In its standard form, the method is not an inference engine. Second, there is no formal separation between observation and intervention. A barrier symbol may indicate a protective function, but it does not automatically make clear whether this function is understood in the model as an observed state, a controllable lever, or both. Third, there is no compact probabilistic state description of the overall system. Qualitative ratings such as \emph{high/medium/low} or \emph{barrier present/not present} are of only limited use for runtime decisions. Fourth, there are no mechanisms for handling dependencies between threats, context factors, and barriers in a principled way. In information and communication technology (ICT) scenarios in particular, common causes, load states, shared observability stacks, or change contexts often invalidate independence assumptions~\cite{khakzad2013,fenton2012}.

These deficits become visible precisely where a risk model is expected to become operationally useful, i.e., in continuous monitoring and rapid intervention. If a material risk must be continuously tracked, the model must be able to distinguish between normal operation, precursors, degradation, and harmful consequences. It must be capable of incorporating observations, updating probabilities, and evaluating possible interventions under current conditions. A Bowtie alone is not designed for this.

\section{Literature Review}

\subsection{Standards}
The described problem is reflected from several perspectives in the literature. A first line of literature is formed by standards and risk management frameworks. ISO~31000~\cite{iso31000} emphasizes that risk management is not a one-time documentation exercise, but a continuous process of analysis, evaluation, treatment, monitoring, and review. IEC~31010~\cite{iec31010} positions Bowtie as an established risk analysis technique, but at the same time clearly separates this communicative and structuring representation from probabilistic modeling and quantitative update mechanisms. In the ICT context, DORA~\cite{dora2022} explicitly requires continuous monitoring, rapid detection, and resilient handling of operational disruptions in the financial sector. NIST SP~800-137~\cite{nist800137} formulates information security continuous monitoring as an ongoing situational awareness process rather than a static assessment procedure. Taken together, this literature clearly points to the need for runtime-capable risk models, but it does not specify how a Bowtie should be transformed into such a model.

\subsection{Bowtie and Bayesian Networks}

A second line of literature concerns the relationship between Bowtie and BN-based modeling. Khakzad et al.~\cite{khakzad2012,khakzad2013} show how Bowtie structures can be mapped into Bayesian networks in order to incorporate dependencies and update probabilities under evidence. Badreddine and Ben Amor~\cite{badreddine2013} also construct Bowtie-like structures within a Bayesian framework. Vairo et al.~\cite{vairo2022} go one step further and reinterpret Bowties as hierarchical Bayesian networks in the context of dynamic risk assessment. Zurheide et al.~\cite{zurheide2021} operationalize the transformation in software with \emph{pyBNBowtie} and thereby show that Bowtie artifacts can, in principle, be transferred to a probabilistic formalism. This work comes close to our own starting point at the level of transformation and software operationalization. A distinct contribution of the present paper is the explicit treatment of the normal state as a runtime modeling requirement and its connection to expert-based probability capture and causal intervention analysis.

The methodological core of the problem lies in the semantics of the normal state. A Bowtie begins at the top event and unfolds causes on the left and consequences on the right. For analysis workshops, this perspective is appropriate because it focuses on the loss path. For monitoring, it is incomplete because the dominant operating case is not the loss, but its absence. The operational state of a system is typically \emph{normal}, \emph{degraded}, \emph{precarious}, or \emph{damaged}; it cannot meaningfully be described only as the right-hand branch of a loss diagram. By contrast, Koller and Friedman~\cite{koller2009} formulate probabilistic graphical models as state spaces with explicit variables, parent relations, and complete local distributions. Fenton and Neil~\cite{fenton2012} likewise emphasize that Bayesian networks are particularly useful when uncertainty, incomplete observations, and expert knowledge are to be combined in a consistent state model. When this perspective is transferred to the Bowtie, it becomes apparent why the \emph{safe state} is not merely a practical addition, but a formal necessity: without an explicit normal state, it remains unclear what evidence in ordinary operation refers to at all.

\subsection{Risk Monitoring}

A third line of literature comes from the field of dynamic risk monitoring. Naderpour et al.~\cite{naderpour2014} develop a situation-awareness approach that understands safety assessment not merely as one-time risk identification, but as the continuous interpretation of a changing situation. Xing et al.~\cite{xing2019} combine condition-monitoring data and inspection data for dynamic risk assessment. Moradi et al.~\cite{moradi2022,moradi2023} combine deep-learning methods with Bayesian networks in order to monitor operating state and risk of complex technical systems simultaneously. Zio~\cite{zio2018} summarizes this development: risk assessment must evolve from static, rarely updated computational pictures to data-rich, dynamic, and decision-proximate procedures. Taken together, these works show that the literature already recognizes the operational necessity of runtime integration. What often remains open, however, is how one can systematically move from a semantically strong risk artifact, such as a Bowtie, to a runtime-capable, evidence-integrable model.

\subsection{Operationalization of Risk Models}

A fourth line of literature concerns Bayesian networks as a bridge between risk analysis and operations. Weber et al.~\cite{weber2012} and Langseth and Portinale~\cite{langseth2007} work out the advantages of Bayesian networks over classic dependability-oriented methods: multivalued states, explicit dependencies, diagnosis and prognosis in a single model, and probabilistic updating under evidence. From a risk-management perspective, Fenton and Neil~\cite{fenton2012} show how Bayesian networks can bring together expert knowledge, data, uncertainty, and decision questions. In their reviews, Kabir and Papadopoulos~\cite{kabir2019} as well as George and Renjith~\cite{george2021} document the increasing shift from static probabilistic risk techniques toward BN-based approaches in safety, security, and process-industry contexts. Cai et al.~\cite{cai2013} demonstrate the applicability of BN-based quantitative risk models to complex operations. In the financial and operational-risk domain, Aquaro et al.~\cite{aquaro2010} and Cornwell et al.~\cite{cornwell2023} show that Bayesian networks can be used not only for analysis after an event has occurred, but for proactive modeling of loss probabilities and causal influence factors. This literature clearly supports the choice of a DAG/Bayesian network as a runtime basis.

However, the same literature also makes clear the cost of this choice: the model structure must be plausible, parametrization is labor-intensive, and the quality of expert judgments is limited. O'Hagan et al.~\cite{ohagan2006} treat the elicitation of expert probabilities as an independent methodological discipline. Werner et al.~\cite{werner2017} show that capturing dependencies in particular is cognitively demanding for experts. Pitchforth and Mengersen~\cite{pitchforth2013} therefore propose validation frameworks for expert-elicited Bayesian networks. Leonelli et al.~\cite{leonelli2023} have recently argued that model checking, sensitivity, and robustness should also be integrated more strongly into toolchains. These works are directly relevant for our contribution because our Realtime Risk Studio not only handles structures, but the integrated Probability Capture tool specifically addresses the difficult parametrization step.

At this point, the work of Kahneman et al.~\cite{kahneman2021} also becomes relevant. Their perspective on noise highlights that judgments can be biased and undesirably variable; the same question can be answered differently by qualified people even though the initial conditions are identical. For probabilistic risk estimation, this means: heterogeneous estimates are not a marginal phenomenon, but a systematic quality characteristic of the elicitation process. For precisely this reason, it is not enough simply to collect expert estimates; they must be analyzed with respect to dispersion, consistency, and prior sensitivity. The noise analysis of our Probability Capture tool is therefore a direct response to a problem described in decision and judgment research.

Taleb~\cite{taleb2007,taleb2010} adds another emphasis. His discussion of black swans and fat tails is relevant for the application field addressed here because realtime risk management is needed precisely for rare, highly impactful, and historically insufficiently observed events. Taleb's objection is directed against epistemic overconfidence in domains in which past data do not provide a reliable, complete representation of future extreme states. For this paper, two consequences follow. First, a runtime model must not be confused with the tacit assumption that rare risks have been sufficiently learned from a few observations. Second, the combination of structured causal modeling, cautious prior setting, and continuous evidence intake becomes especially valuable in highly uncertain domains.

Finally, a distinction must be made between \emph{observational} and \emph{interventional} uses of the model. Pearl~\cite{pearl2009} distinguishes among association, intervention, and counterfactuals and provides a formal language, named do-calculus, for when causal effects may be identified from a graph. Huntington-Klein~\cite{huntingtonklein2021} formulates this idea very pragmatically in modern textbook form as a design question: which paths must be closed, which variation must be isolated, and which control decisions must be made so that a statement can be interpreted causally. This distinction is central for realtime risk management. Monitoring alone answers only how likely a risk is. For timely response, however, one additionally needs to know which intervention at which node is likely to improve the situation.

In summary, the literature thus shows a consistent picture: standards require continuous observation; Bowties structure risks very well, but are limited as direct runtime models; Bayesian networks are suitable for evidence-based updating; causal models in the sense of Pearl are relevant for intervention questions; elicitation, noise, and model robustness remain methodologically demanding. This Part~3 of the series addresses this gap by integrating methods and tools that (i) systematically transform a Bowtie into a runtime-capable DAG, (ii) explicitly model the \emph{safe state}, (iii) capture probabilities in a structure-based and subgraph-specific way, (iv) treat noise and priors explicitly, and (v) enable causal analyses for later decisions on the same structure.

\section{Related Work}

\subsection{DAGs for Representing States and Causal Relationships}
A DAG is a directed graph without cycles. In probabilistic graphical models, the nodes represent random variables and the edges represent direct dependencies; acyclicity ensures that a topological ordering and thus a well-defined factorization of the joint distribution exist~\cite{pearl1988,koller2009,darwiche2009}. In the classical BN view, a DAG factorizes the joint distribution as
\begin{equation}
P(X_1,\dots,X_n)=\prod_{i=1}^{n} P(X_i\mid Pa(X_i)),
\end{equation}
where $Pa(X_i)$ denotes the parent set of $X_i$. It is precisely this local factorization that makes DAGs attractive for risk models: global uncertainty is decomposed into local probability assumptions that can be formulated in domain language.

Koller and Friedman~\cite{koller2009} systematize DAGs as the core structure of probabilistic graphical models and show how independence assumptions, inference, and learning build on this structure. Darwiche~\cite{darwiche2009} complements this view with a strong focus on modeling, inference, and sensitivity, making clear that Bayesian networks are not only compact representations, but operational reasoning models. The algorithmic foundations of efficient inference were shaped early on, among others, by Lauritzen and Spiegelhalter~\cite{lauritzen1988}. For the present paper, the key point is this: a DAG is not merely a pretty visualization, but the smallest formal unit in which structure, uncertainty, and evidence updating coincide.

This point fundamentally distinguishes a DAG from many risk artifacts used in practice. A Bowtie is semantically strong, but it is typically not specified as a complete factorization of a state distribution. A DAG, by contrast, forces every node to have a clear state space and explicit parent relationships. This also makes clear what is missing. For our problem, this is the \emph{safe state}: as long as the normal state is not modeled as a regular state of a target or consequence node, it remains unclear how sensing and evidence should be interpreted during non-damaging operation.


\subsection{Causal Analysis}
Pearl~\cite{pearl1988,pearl2009} has decisively shaped the transition from directed graphs as mere probability models to directed graphs as \emph{causal} models. For our context, the distinction between observation and intervention is particularly central. A monitoring model that provides only associations or posteriors under evidence does not yet answer the question of which measure is causally effective. Do-calculus, backdoor and frontdoor criteria, and the formal distinction between observing and setting are precisely what make it possible to move from \enquote{recognizing the current risk state} to \enquote{selecting an appropriate response}~\cite{pearl2009}.

The causal DAG literature has further operationalized this core. Spirtes et al.~\cite{spirtes2000} developed a foundational bridge between graphs, conditional independence, and causal discovery. Spirtes and Zhang~\cite{spirtes2016} provide a compact overview of modern concepts and methods of automated causal inference and discovery. Glymour et al.~\cite{glymour2019} systematically classify constraint-based, score-based, and functional approaches. Textor et al.~\cite{textor2016} bring this theory into a practical form with \texttt{dagitty}, especially for adjustment sets and bias-path analysis. Huntington-Klein~\cite{huntingtonklein2021} complements this mathematical tradition with a particularly clear, design-oriented explanation of what \enquote{identification} means in practice: making paths visible, closing unwanted backdoors, and isolating the variation that may actually be interpreted causally.

Our Realtime Risk Studio is intended not only to display probabilities of states, but, after parametrization, also to support causal-analysis questions such as \emph{minimal adjustment sets}, \emph{backdoor/frontdoor adjustment}, \emph{d-connected trails}, \emph{node independencies}, \emph{node inference}, and impact-oriented do-analyses. These concepts do not come from the Bowtie world, but from the causal analysis world. Their integration into a tool is therefore methodologically relevant: it transforms a risk artifact into a decision artifact.

\subsection{Tools and Software Ecosystems}
The methodological ecosystem around DAGs, Bayesian networks, and causal analysis is broad, but fragmented. For graphical modeling and inference, GeNIe/SMILE~\cite{druzdzel1999} and HUGIN~\cite{madsen2003} are among the established tools. Both support the creation, parametrization, and evaluation of probabilistic models, but they do not directly address the Bowtie-to-DAG operationalization including domain-specific Probability Capture workflows that is central to this paper.

For statistical and programmatic work with Bayesian networks, \texttt{bnlearn}~\cite{scutari2010}, \texttt{pgmpy}~\cite{ankan2024}, and \texttt{bnmonitor}~\cite{leonelli2023} are particularly relevant. \texttt{bnlearn} is strong in structure learning and parameter estimation from data, \texttt{pgmpy} covers a wide spectrum of BN and causality functions in Python, and \texttt{bnmonitor} addresses model checking, sensitivity, and robustness. These tools are methodologically strong, but they typically presuppose an already formalized graph model.

For graph-based causal analysis, \texttt{dagitty}~\cite{textor2016}, \texttt{pcalg}~\cite{kalisch2012}, and \texttt{causaleffect}~\cite{tikka2017} are notable. \texttt{dagitty} is strong in graphical inspection, adjustment sets, and bias-path analysis; \texttt{pcalg} supports discovery and effect-estimation procedures; \texttt{causaleffect} operationalizes identification using do-calculus. For causal discovery in Python, \texttt{causal-learn}~\cite{zheng2024}, the \texttt{Causal Discovery Toolbox}~\cite{kalainathan2020}, and the \texttt{LiNGAM} package~\cite{ikeuchi2023} are particularly relevant. Together, these packages form a strong ecosystem, but none of them is designed as a toolchain for realtime risk management starting from Bowtie artifacts.

This is precisely where our approach differs. The Realtime Risk Studio does not begin with an abstract statistical problem and not with an already finished causal graph, but with a Bowtie-structured risk artifact. It adds the missing \emph{safe state}, maps barriers into observable activation and state nodes, generates a DAG from them, supports subgraph-specific Probability Capture workflows, makes noise explicitly visible, and subsequently provides causal analysis functions on the same structure. The existing literature and tools thus provide essential building blocks, but not the integrated bridge we need from the risk workshop to the realtime monitoring and analysis model.

\section{Hagenberg Risk Management Process}
So far, Part~1~\cite{hermann2026part1} of the Hagenberg Risk Management Process addresses the problem that classic two-dimensional heatmaps inadequately capture context dependencies of complex systems. The paper introduces multidimensional polar heatmaps in which additional context axes (e.g., redundancy level, maintenance status, load states, or environmental conditions) are modeled explicitly. Methodologically, this is important because risks in modern infrastructures are rarely invariant. The same risk can have a different character, e.g., under peak load, during change windows, or under degraded third-party conditions than it does in normal operation. Part~1 thus creates the triage logic of the overall process: which risks are materialized under which contexts and therefore deserve deeper analysis?

Part~2~\cite{hermann2026part2} translates this triage logic into a semantic case analysis. Context-sensitive risks are transformed into Bowties so that causes, top event, consequences, and barriers are systematically captured. It is central that barriers are not only documented, but treated as their own semantic units. Part~2 further shows how Bowtie structures can be transformed into a DAG, which in turn serves as the structural basis of a Bayesian network. A particular contribution is the introduction of \emph{Activation Nodes}: barriers are modeled in such a way that they can later be not only observed in the system, but also recognized as potential intervention points.

Part~3 starts exactly at the unresolved points of Part~2. There, probabilistic parametrization, evidence integration, intervention analysis, and continuous monitoring were explicitly named as future work~\cite{hermann2026part2}. The present paper closes this gap on three levels.

First, it operationalizes the Bowtie-to-DAG idea through an explicit \emph{safe state}. This formally completes the monitoring concept already implied in Part~2. Second, with the Probability Capture tool, it provides a method for systematically parameterizing the CPTs of the resulting Bayesian network by means of expert estimates and optional priors. Third, a causal-analysis mode extends the activation concept known from Part~2 toward do-analyses, adjustment sets, d-connected trails, and evidence-dependent intervention search.

In other words, this yields an end-to-end process: Part~1 answers \emph{which} risks should be examined in greater depth and under which contexts. Part~2 answers \emph{how} a prioritized risk is semantically modeled and transformed into a probabilistic structure. Part~3 finally answers \emph{how} this structure is parametrized, causally analyzed, and prepared for realtime monitoring and decision support. Taken together, the three contributions thus do not form a loose toolbox, but a consistent pipeline from governance-capable overview to an operationally connectable runtime model.

\section{Modeling a Realtime Risk Model}

At the center of Part~3 is the question of how a runtime-capable risk model emerges from a semantically structured Bowtie. This model must possess three properties simultaneously. First, it must preserve the risk logic developed in Part~2. Second, it must provide an explicit state space for observation and inference. Third, it must represent interventions in such a way that later measure analyses become causally meaningful. Our Realtime Risk Studio serves as a reference implementation in which these modeling steps are explained concretely.

The methodological focus therefore lies on the \emph{modeling of the realtime risk model}. The crucial point is that modeling does not begin only at the monitoring stage. Already when a DAG is being built, it is determined which states can later be observed, which relationships can be evaluated probabilistically, and which protective functions can be treated as intervention-capable nodes.

\subsection{Bowtie $\rightarrow$ DAG as a Structural Transformation}
The Bowtie-to-DAG mapping is not merely a redrawing of the same contents with arrows, but a semantic structural transformation. The left side of the Bowtie provides the preventive logic: causes, preconditions, gates, and barriers determine whether the top event occurs. The right side provides the mitigative logic: after the top event occurs, alternative event and consequence paths unfold, whose course in turn depends on barriers and context conditions. In a DAG, these two views are transformed into a unified causal structure.~\cite{zurheide2021}

\begin{table}[t]
\caption{Mapping rules Bowtie $\rightarrow$ DAG}
\label{tab:mapping}
\centering
\footnotesize
\begin{tabular}{p{2.3cm} p{4.9cm}}
\toprule
Bowtie element & DAG representation \\
\midrule
Threat / cause & Node with discrete states; directed edges to intermediate events or the top event \\
Gate (AND/OR) & Deterministic node with correspondingly generated CPTs \\
Top event & Central event node connecting preventive and mitigative subgraphs \\
Barrier & Node with states such as \emph{works}/\emph{fails}; can additionally be marked as an Activation Node \\
Consequence paths & Merged into outcome or consequence nodes with alternative end states \\
Normal operation & Additional state \emph{safe}, representing the case \emph{Top Event = false} \\
Context and preconditions & Additional parent nodes modeling load, operating mode, resource situation, or trigger conditions \\
Runtime integration & Enrichment of individual nodes with REST sources and notify targets for later evidence integration \\
\bottomrule
\end{tabular}
\end{table}

Tab.~\ref{tab:mapping} summarizes the essential mapping rules of the present approach.
The decisive difference compared with merely transferring structure lies in the outcome space. Consequences are not only represented as several loose paths to the right of the top event. The model brings together the possible end states in an explicit consequence space and adds the \emph{safe} state. Thus, for a consequence node $C$ with state set $\mathcal{S}(C)$,
\begin{equation}
\sum_{s \in \mathcal{S}(C)} P(C=s \mid Pa(C)=u)=1
\end{equation}
holds for every parent configuration $u$. The state \emph{safe} thereby represents normal operation, i.e., precisely the case in which no loss-relevant path has occurred. Only this makes the model useful for monitoring as explained in previous sections.

\subsection{Contextual Embedding of Bowtie Nodes}
A central added value of the modeling style proposed here is that Bowtie nodes can be \emph{contextually embedded} within a DAG. A Bowtie describes in compact form which threats, barriers, events, and consequences are relevant for a risk. For realtime monitoring, however, this condensation is often not sufficient. Additional causal parent nodes can therefore be introduced in the DAG in order to model \emph{under which preconditions} a Bowtie node occurs, succeeds, or is effective in a specific form.

This is particularly important for barriers. A Bowtie often records a barrier only as an existing protective function. In a realtime risk model, however, we also model the conditions on which its success depends. A barrier such as \emph{Automatic Rollback} may, for example, depend on the preconditions \emph{Rollback Pipeline Available}, \emph{Observability Healthy}, and \emph{Change Scope Compatible}. In the DAG, the barrier is thus represented not only as a node with states \{\emph{works}, \emph{fails}\}, but can be modeled context-sensitively, for example with states such as \{\emph{works}, \emph{works-delayed}, \emph{fails}\}. In this way, the model does not merely capture {whether} a protective function works, but also {how} it works under different preconditions and which consequence paths thereby remain open.

The same applies to event nodes taken over from the event tree (Bowtie right side). In a Bowtie, such an event often appears as a simple branch. In a DAG, it can be modeled with additional parents so that an event trigger is tied to concrete conditions. An event such as \emph{Retry Storm} may, for example, depend on \emph{High Latency}, \emph{Client Retry Discipline}, and \emph{Backoff Policy}. At the same time, its state space can be refined, for example to \{\emph{none}, \emph{local}, \emph{sustained}\}. This makes it possible to express that the event does not merely occur or not occur, but can occur in different forms that affect the consequences to different degrees.

This contextual embedding therefore does not arbitrarily extend Bowtie semantics, but operationalizes them. Bowtie nodes remain the semantic anchors of the risk model, but in a DAG they are complemented by those preconditions that are actually relevant for observation, evidence integration, and intervention. It is exactly this that creates a model capable of distinguishing between \enquote{barrier exists} and \enquote{barrier works under the currently observed conditions with a specific quality}.

\subsection{Modeling Workflow in the Reference Implementation}
The reference implementation supports this modeling workflow in a graphical environment. Nodes can be created, deleted, renamed, and equipped with user-defined state spaces; edges can be added or removed; subsets of nodes can be copied, duplicated, and further processed in subgraphs. 

In practice, modeling begins with a Bowtie. Threats, gates, top event, barriers, and consequence paths are transferred into a DAG~\cite{zurheide2021}. It is then checked which of these nodes are still too coarse for the runtime model. Exactly at this point, additional parents are introduced: load and context indicators, operational preconditions, technical dependencies, or organizational prerequisites. The state spaces are then refined. Where a binary state is sufficient in domain terms, the model remains compact; where different manifestations are relevant for monitoring or decisions, multivalued states are used.

An important design principle here is the deliberate limitation of the number of parents per node. In Bayesian networks, the conditional probability table (CPT) grows exponentially with every additional parent variable in the worst case. The reference implementation therefore closely couples structure modeling and parameterizability: additional parents should be introduced only where they either significantly clarify the causal structure or are actually needed for the later evidence and intervention logic. This prevents both semantic overloading and CPTs that are impractical to elicit.

The model is then transferred into an XML-based representation. This intermediate representation is methodologically important because it standardizes the transition from visual modeling to machine-readable artifacts. From this structure, the Probability Capture tool questions, the CPT representations, the monitoring integration, and the causal-analysis functions are derived. The reference implementation additionally provides a raw-structure editor for the exported representation so that modelers can inspect, control, and, when necessary, adjust the structural export directly.

\subsection{Activation Nodes for an Intervention-Capable Model}
Part~2 introduced barriers as \emph{Activation Nodes}~\cite{hermann2026part2}. Activation Nodes are nodes that can not only be observed, but can also be deliberately set during operation. These may be classic barriers such as \emph{Automatic Rollback}, \emph{Traffic Shedding}, or \emph{Regional Isolation}, but also organizational-technical measure nodes.

The separation between observation and intervention is central. A node with an external source, e.g., a REST interface, initially stands for evidence: it states which state \emph{is}. An Activation Node, by contrast, stands for a deliberately induced measure: it represents which state \emph{can be set}. This distinction is methodologically necessary for later do-analyses in Pearl's sense, because only then can observational probability and intervention effect be cleanly distinguished~\cite{pearl2009}. The contextual enrichment from the previous subsection is retained: intervention-capable nodes may also have parents for modeling under which conditions a measure is likely to be fully effective, delayed, or only partially effective.

\subsection{Probability Capture}
After structural modeling comes the probabilistic enrichment of the model. The Probability Capture tool generates the necessary questions directly from a DAG. For each row of a CPT, that is, for each concrete parent-state configuration, questions are generated in such a way that all target states except the last are asked explicitly. This design decision uses the normalization condition of the CPT: if the sum of all state probabilities is~1, the last state can be computed from the others. This reduces the number of values to be elicited without loss of information, but it also imposes an important validation requirement: the elicited probabilities for the first $K-1$ states must not sum to more than~1. If they do, the row is invalid and must be revised rather than silently accepted.

The generated questions are semantically formulated as conditionals, in the style of \enquote{What is the probability that node~X has state~s, given that parent nodes~A and~B are in states~a and~b?} This supports not only numerical elicitation, but also the \emph{anchoring} of the questions: time horizon, scope, state definition, and conditioning are to be formulated precisely and comparably for all experts.

Another advantage of the modeling style is that questions are generated not only globally for the overall model, but also for subgraphs or selected node regions. This makes it possible to target exactly those domain experts who actually possess robust knowledge about the respective context. For example, deployment experts are more likely to answer questions about \emph{Validation Gate} or \emph{Automatic Rollback}; site reliability engineers are more likely to answer questions about \emph{Queue Saturation}, \emph{High Latency}, or \emph{Retry Storm}; business or operations-risk roles are more likely to answer questions about the manifestation of consequence states. The reference implementation supports this through selective capture scopes and domain-tailored question packages.

In addition, the tool allows quick-set values such as \emph{None}, \emph{Very low}, \emph{Low}, \emph{Medium}, \emph{High}, \emph{Very high}, and \emph{Evidence}. These scale anchors reduce friction in the capture process without giving up precise numerical storage. Furthermore, priors can be stored per question: an expected baseline value $p_0$ and a prior strength $k_{prior}$. In methodological terms, these parameters are best understood as a way to introduce domain-informed regularization when responses are sparse, highly dispersed, or intentionally anchored to known base-rate expectations. The reference implementation also includes optional artificial intelligence (AI) support in which project-specific documents are used as context for assisted elicitation. This is not essential to the scientific core contribution, but it shows the compatibility with assisted modeling processes.

\subsection{CPT Estimators and Noise Analysis}

A fundamental problem in obtaining probabilities from expert estimates is noise, that is, unwanted dispersion of judgments about the same matter~\cite{kahneman2021}. For the Probability Capture tool, it is not crucial to diagnose every psychological subcategory separately, but to make the resulting dispersion across repeated or parallel estimates per question visible and to separate this from the actual aggregation of probabilities.

Suppose $n$ experts have been asked the same question, e.g., \enquote{What is the probability of $X$ given $Y$?} and the answers are given in chronological order as $y_1,\dots,y_n \in [0,1]$. CPT estimates and noise analysis are computed for every question based on these responses. Furthermore, the computation assumes global parameters that can also be overridden per question, i.e., starting probability $p_0$, base weight $k_{prior}$, expert concentration~$\kappa$, and a half-life parameter. In the reference implementation, these settings are modifiable in a web form,  stored persistently, and reused in diagnostics and  final CPT materialization. 

We propose three different methods for analyzing dispersion of answers: \emph{Equal Average}, \emph{Anchored Average}, and \emph{Expert Consensus}.
Furthermore, we define seven estimators for aggregating $y_1,\dots,y_n$ into a single CPT probability: \emph{Equal Average}, \emph{Middle Value}, \emph{Balanced Average}, \emph{Anchored Average}, \emph{Expert Consensus}, \emph{Latest Answer}, and \emph{Cautious Average}.
The coincidence in naming of the methods is on purpose because estimation and noise analysis are connected in those cases.
Figure~\ref{fig:noise-analysis} illustrates the noise analysis in Probability Capture tool, where estimates are enumerated and dispersion is shown in three spread bars in the probability space.

\subsubsection{Equal Average}
The simplest estimator is the arithmetic mean of the expert responses with

\begin{equation}
\hat p_{M}=\frac{1}{n}\sum_{i=1}^{n} y_i
\label{eq:mean}
\end{equation}

as the central descriptive location measure. The advantage of this method lies in its simplicity. It is easy to communicate, has no hyperparameters, requires no additional probabilistic assumptions, and remains transparent because every submitted value contributes linearly and with equal force. The disadvantage is that the estimator takes into account neither an explicit base rate nor a model of heterogeneous expert precision. A few extreme entries can noticeably shift the value, and with very small sample sizes per question, a resulting CPT entry is often unstable. For operational applications, the average is useful as a baseline, but not a robust standard estimator when data are sparse. 

For noise analysis based on the arithmetic mean, residuals are defined as
\begin{equation}
 r_i = y_i - \hat p_{M},
\end{equation}
and, for $n>1$, the unbiased sample variance is
\begin{equation}
 s^2 = \frac{1}{n-1}\sum_{i=1}^{n} r_i^2,
\end{equation}
with standard deviation $s=\sqrt{s^2}$. The spread bar in the user interface then draws $\hat p_{M}\pm\sigma$, clamped to the interval $[0, 1]$ to show the dispersion of expert responses.


\subsubsection{Middle Value and Balanced Average}
The aforementioned methods are two weighted summaries for CPT estimation. The \emph{Middle Value} is the weighted median that returns the 50\% quantile of the elicited probabilities and is therefore more robust to isolated high or low outliers than the arithmetic mean. The \emph{Balanced Average} is the weighted logit mean. The expert responses are transformed to log-odds space, a weighted average is computed using the half-life decay weights of the answers, and the result is transformed back to the probability interval $[0,1]$. This tends to behave more symmetrically near the boundaries 0 and 1 than direct averaging on the probability scale.

Methodologically, both variants occupy a middle ground with respect to estimation. They are less assumption-heavy than the \emph{Anchored Average} and \emph{Expert Consensus} estimators, but they are more structured than an arithmetic mean. In practice, the median is attractive when robustness against outliers is desired, whereas the logit mean is useful when probability compression near the boundaries should be reduced without moving to a full latent-consensus model.

\subsubsection{Anchored Average}
This estimator implements a prior-regularized weighted mean. For each question, the tool stores a prior mean $p_0$ and a prior strength parameter $k_{prior}\ge 0$. Let
\begin{equation}
a_0=p_0 k_{prior} \quad \text{with } a_0 > 0,
\end{equation}
\begin{equation}
b_0=(1-p_0)k_{prior}\quad \text{with } b_0 > 0,
\end{equation}
with expert responses $y_1,\dots,y_n$ and, if enabled, half-life-based weights $w_i\ge 0$,
\begin{equation}
S=\sum_{i=1}^{n} w_i y_i, \qquad T=\sum_{i=1}^{n} w_i,
\end{equation}
the computed estimate is
\begin{equation}
\hat p_{A}=\frac{a_0+S}{a_0+b_0+T}.
\end{equation}

With $k_{prior}=0$ the prior is deactivated and the estimator reduces to the weighted mean; without half-life weighting, $w_i=1$ for all answers.

For noise analysis, the 95\% credible interval is the equal-tail interval of the implied Beta$(a_0+S,b_0+T)$ distribution.
This method provides stabilization for sparse or highly dispersed expert responses. The method is \enquote{anchored} because the prior $p_0$ signifies a domain-justified base rate, while $k_{prior}$ controls the strength of this anchoring relative to the observed judgments from experts. For questions with only few responses, this often yields more stable estimates than an arithmetic mean.
Algebraically, the formula resembles a posterior mean from a Beta-Binomial model, but it should be interpreted as a pragmatic shrinkage rule for point-probability elicitation rather than a literal Bayesian update from binary trial data.

\subsubsection{Expert Consensus}
To aggregate continuous expert probability judgments $y_1,\dots,y_n$, we employ a latent-consensus model in which all elicited values are interpreted as noisy assessments of a common underlying probability $p$. When prior information is enabled ($k_{prior} > 0$), the latent consensus is assigned a Beta prior,
\begin{equation}
p \sim \mathrm{Beta}(a_0,b_0), \quad a_0 = p_0 k_{prior}, \quad b_0 = (1-p_0)k_{prior},
\end{equation}
where $p_0$ denotes the prior mean (i.e., domain knowledge) and $k_{prior}\ge 0$ controls prior strength. Conditional on $p$, each expert judgment is modeled as
\begin{equation}
y_i \mid p \sim \mathrm{Beta}(\kappa p,\kappa(1-p)),
\end{equation}
so that $\mathbb{E}[y_i\mid p]=p$, while the expert concentration parameter $\kappa>0$ governs how tightly individual judgments cluster around the latent consensus. The default value in the reference implementation is $\kappa = 8.0$. Optional half-life weighting introduces recency-sensitive weights $w_i$, leading to the weighted posterior kernel
\begin{equation}
\pi(p\mid y_1,\dots,y_n)\propto \pi(p)\prod_{i=1}^n f(y_i\mid p)^{w_i}.
\end{equation}
As the resulting posterior is not available in closed form, it is approximated numerically on an adaptive two-stage grid: a coarse grid (121 points) is first used to locate the region of highest posterior mass, then a finer grid (401 points) is placed around this region to obtain precise summary statistics. The quantity entered into the CPT is the posterior mean $\hat p_{EC}$, along with the posterior standard deviation and 95\% equal-tail credible interval for noise analysis. If no elicited probabilities from experts are available, the routine returns the prior mean; with $k_{prior}=0$ (no prior), a uniform reference $Beta(1,1)$ is returned.

Compared with \emph{Anchored Average}, both methods use a Beta-distributed prior, the \emph{Expert Consensus} treats expert judgments as noisy observations of a latent consensus probability, whereas \emph{Anchored Average} directly combines the prior with weighted point estimates. The parameter $\kappa$ controls the assumed observation noise, and the posterior is obtained numerically rather than analytically. 

In statistical modeling approaches, future developments are typically inferred from historical data, implying an extrapolation of past patterns into the future. Such an approach is inherently limited, as previously unknown yet plausible events cannot be captured by the underlying data-generating process. To address this limitation, an expert-consensus framework can be employed that combines historical evidence, represented by \(k_{\mathrm{prior}}\) and a starting probability $p_0$, with expert judgments on expected future developments. In addition, the introduction of a half-life parameter allows the relative influence of forward-looking expert knowledge to increase progressively over time, thereby enabling a gradual shift in emphasis from historical information toward future expectations.

\subsubsection{Latest Answer}
As the simplest operational rule, the tool can also directly adopt the most recently entered response for a question:
\begin{equation}
\hat p_{L}=y_n.
\end{equation}
This mode does not involve statistical estimation in the narrower sense, but merely reflects the latest available input state.
The advantage of this method is its simplicity in iterative workshops or review situations. If, for example, only the most recent revision is deliberately meant to apply, such as after an explicit discussion or a targeted correction round, the \enquote{Latest Answer} mode is an understandable operational mechanism.

The disadvantage is obvious: earlier expert estimates, dispersion, and prior information are ignored. For robust group aggregation, this method is therefore inapplicable and should not be understood as an estimator. It should be read more as a convenience and workflow function than as a statistically preferred estimation method.

\subsubsection{Cautious Average}
As an alternative aggregation rule, the tool provides the root mean squared (RMS) estimator:
\begin{equation}
\hat p_{RMS}=\sqrt{\frac{1}{n}\sum_{i=1}^{n} y_i^2},
\end{equation}
where $y_i$ are the elicited expert judgments. By squaring the individual values, high probabilities are weighted more strongly than in the arithmetic mean. The RMS therefore does not represent a neutral group center, but a deliberately high-value-emphasizing compression of expert estimates.
Unlike the other estimators in the tool (Equal Average, Anchored Average, Expert Consensus), the RMS does not apply half-life weighting; it operates directly on the raw elicited values without temporal discounting.
The advantage of this method is that high individual judgments are not diluted as strongly. In situations in which the model is intended to aggregate more cautiously or in a more alarm-sensitive fashion, this may be desired. The RMS can therefore be meaningful as a policy rule if high-risk assessments of individual experts are deliberately intended to have a stronger impact on the CPT.
The disadvantage is that RMS is not a standard aggregation for probabilities. For dispersed responses, it lies systematically above the arithmetic mean and can therefore shift probabilities upward without constituting a statistical bias correction. For the noise analysis itself, the implementation therefore does not use the RMS as the central location measure, but only as an optional alternative aggregation rule or additional metric.

\subsubsection{Remarks}

Each estimation function in the Probability Capture tool serves a different purpose. \emph{Equal Average} is the transparent baseline, \emph{Middle Value} is the robust median-style summary, \emph{Balanced Average} is a boundary-aware compromise on the log-odds scale, \emph{Cautious Average} is a deliberately high-value-emphasizing special rule that is best interpreted as a conservative policy heuristic rather than as a neutral statistical center, \emph{Latest Answer} is an operational special function for the latest input state, the prior-regularized \emph{Anchored Average} is the base-rate-sensitive regularized update, and \emph{Expert Consensus} is the most structured bounded-support aggregation of several point judgments. In practical terms, the first five should primarily be read as diagnostic or policy-oriented summaries, whereas \emph{Anchored Average} and \emph{Expert Consensus} are the model-based estimators most naturally suited for computing CPT entries when sparse data or disagreement need to be handled explicitly. This separation is important in practice: dispersion is made visible separately, while the actual estimator can be chosen according to domain objective and data situation.

\subsection{Causal Analysis on the Completed Model}
Once DAG modeling has been completed and the CPTs have been filled, the model is no longer used primarily as a structural diagram, but as a runtime model that supports causal what-if reasoning under the assumed causal structure. Methodologically, this means that the DAG now fulfills three roles simultaneously. First, it represents assumptions about directed influence relationships among causes, amplifiers, barriers, and consequences. Second, it allows probabilistic inference under evidence. Third, it creates the prerequisite for distinguishing hypothetical interventions from observed evidence. In this context, the implementation of Realtime Risk Studio serves primarily as a reference for which of these methodological operations are supported.

\subsubsection{Adjustment and identifiability} A central step in any causal analysis is to clarify under which conditions an effect of interest is interpretable at all. For an exposure $X$ and a target $Y$, it is not sufficient that the two are connected nodes in the graph. It must be determined which third variables act as confounders and therefore need to be controlled. Adjustment sets serve precisely this purpose. A minimal adjustment set denotes the smallest set of variables whose conditioning suffices to ensure that the causal effect of $X$ on $Y$ is not distorted by common causes. Methodologically, this is especially important for the risk model because operational relationships often contain several context factors acting in parallel: a precondition or an already active barrier can explain an apparently direct relationship between two nodes even though, in truth, a confounded path is present. Determining minimal backdoor or, in special structures, frontdoor-based control sets is therefore not merely a technical convenience feature, but a means of obtaining a disciplined causal reading from the completed DAG.

\subsubsection{D-separation as a structural check} Before effects are interpreted or interventions compared, it should be checked whether the independence assumptions encoded in the model are plausible in domain terms. In DAGs, this can be formulated via d-separation. Two nodes are d-separated under a given evidence set exactly when all relevant paths between them are blocked; they are d-connected if at least one active path remains open. This check is central for model quality. It forces the translation of Bowtie or domain knowledge into precise semantic statements: which paths should be open under normal assumptions, which only under certain observations, and which should be blocked in principle? Particularly in more complex risk models, this is an important review technique. A path that should be considered blocked in domain terms but remains d-connected in the DAG typically indicates a missing context variable, a misdirected edge, or an insufficient barrier structure.

\subsubsection{Probabilistic inference under evidence} On a fully parametrized Bayesian network, observation-based inference can then be carried out. In this process, evidence is set on selected nodes of a model, and an algorithm computes posterior probabilities for other node states. Methodologically, this operation answers questions of the form: \enquote{How does the risk situation change if we know that certain preconditions, observations, or warning signals are present?} For realtime risk management, this is the bridge from a static model to ongoing situational assessment. Observations, e.g., \emph{high latency} in the case study in the next section, are not interpreted in isolation, but embedded into the overall hypothesis of the model. The result is not a binary diagnosis, but an updated state distribution from which one can read which consequences become more plausible under the current evidence and which protection assumptions still hold.

\subsubsection{Intervention instead of observation} The methodologically decisive added value of causal models, however, lies in the fact that observation and intervention can be treated separately. Assigning evidence to a node does not automatically correspond to an intervention in Pearl's sense~\cite{pearl1988,pearl2009}. If it is observed, for example, that a barrier is active, this initially says only something about the state of the system under the existing causes. An intervention of the form $do(X=x)$, by contrast, cuts the incoming edges into $X$ and asks how the system would behave if this state were actively brought about. In the Realtime Risk Studio, this distinction is particularly important for Activation Nodes. Measures, e.g., \emph{automatic rollback} in the case study, are not only observable states, but genuine intervention candidates. Causal analysis on the completed model therefore means comparing the hypothetical effect of such interventions under the given evidence and boundary conditions, rather than hastily inferring effectiveness from mere correlations.

\subsubsection{Comparing measures under situation dependence} Building on this do-semantic view, interventions can be evaluated not only individually, but also weighed against one another. Methodologically, the relevant questions are then: which measure reduces the probability of a harmful state most strongly under the current evidence? And at which point can the largest marginal leverage on a target state be expected? Such comparisons are especially relevant in realtime risk scenarios because the best intervention is rarely absolute, but depends on context. For example, in the case study, a rollback may be highly effective in one scenario, yet already too late in another; traffic shedding may prevent a complete outage under certain conditions, but under others may only limit the severity of damage. The implementation supports this kind of search for effective intervention points, but methodologically it should be understood more generally: the completed DAG becomes an instrument for systematically exploring the intervention space rather than merely describing symptoms.

\subsubsection{From model checking to decision support} To sum up, once parametrization of a DAG has been completed, the focus shifts from construction to using it as a formal medium of argumentation and decision support. Under the assumed model, adjustment sets support interpretable effect estimation, d-separation checks the semantic consistency of the structure, probabilistic inference updates the situation under evidence, and do-based analyses compare hypothetical actions. The reference implementation operationalizes these steps, but methodologically its value lies in the fact that a model does not stop at a plausible visualization. It becomes an explicit representation of which causal assumptions are being made, how evidence enters into situational assessment, and which interventions can be provisionally preferred under those assumptions.

\subsection{Foundation for Realtime Monitoring}

Modeling does not end with analytical results. Already during the construction of a DAG, nodes are prepared for later runtime integration in the reference implementation. RESTful interfaces for event sources and notification targets can be stored per node; in addition, the system can generate REST endpoints that provide node status or forward state changes for realtime monitoring. This refinement shows that a DAG is not conceived as a static research artifact, but as the basis of evidence-driven situational awareness.

For monitoring, the coupling of explicit state space, evidence integration, and causal structure is especially important. Nodes with observable states can be fed by telemetry, APIs, alerts, or external models. Any change in evidence propagates so that posteriors for states and consequences are updated in all nodes. Activation Nodes remain deliberately separated from evidence because they represent measures to be taken.
In this way, a clean separation between \emph{perceiving} and \emph{intervening} can be achieved.
Our reference implementation, the \enquote{Monitoring and Decision Hub}, will be the focus of Part~4 in this publication series.

\section{Case Study: Instant-Payments Gateway}
\subsection{Initial Situation and System Boundary}
The instant-payments gateway introduced in Part~2 is used as a running example~\cite{hermann2026part2}. The case is DORA-related and realistic: a financial infrastructure operates a critical payment service across active/active regions or availability zones. The gateway receives payment orders, validates and transforms messages, writes them into internal processing and persistence paths, communicates with downstream services, and returns a response to the invoking channel or connected system within tight time constraints. Precisely because instant payments are time critical, a technical degradation quickly becomes a business-critical problem: even a few additional seconds of delay can lead to retries, operational escalations, and, in the extreme, lost transactions or transactions that can no longer be reconstructed unambiguously.

For the case study, the system boundary is deliberately chosen so that it captures the operational core problem. Not all conceivable business and compliance aspects of payment processing are considered, but those components and states that are relevant for runtime observation of the material risk: production change, rollout mechanics, load situation, queue behavior, latency, retry behavior, regional isolation or traffic steering, rollback, and the resulting manifestation of damage. This restriction is methodologically important. A realtime risk model is not intended to represent the entire reality of the enterprise, but rather that substructure in which observable evidence, relevant barriers, and meaningful interventions actually come together.

In the example, the disturbance begins with a faulty production change, for example, an unfavorable routing rule, a faulty timeout configuration, or a parameter that remains unnoticeable under normal operation but becomes problematic under load. The load window is not merely a background factor, but a causal amplifier: under high transaction rates, queues build up more quickly, congestion spreads further, latency artifacts become more difficult to interpret, and protective mechanisms such as auto-rollback, traffic shedding, or regional isolation reach their limits under different boundary conditions than in idle mode or during tests.

\subsection{Concrete Bowtie Structure of the Example}
For the Bowtie, the left, preventive part can be structured relatively clearly. In particular, \emph{faulty production change}, \emph{routing misconfiguration}, \emph{insufficient rollout validation}, and \emph{peak operational load} appear as threats or causal paths. Preventive barriers include, among others, \emph{config/routing validation}, \emph{pre-deployment tests}, \emph{change approval}, and \emph{canary release with automated rollback}. The top event is modeled as \emph{service degradation under load}. This formulation is deliberate: the top event is not already \enquote{complete failure}, but the operational tipping point at which the service can no longer reliably maintain its specified quality under load.

On the right side of the Bowtie, the consequence paths are further differentiated in domain terms. One plausible sequence is, for example, \emph{high latency} $\rightarrow$ \emph{retry storm} $\rightarrow$ \emph{partial unavailability} $\rightarrow$ \emph{complete unavailability} $\rightarrow$ \emph{lost transactions}. However, there are also alternative courses, for example a situation in which latency is high and the queue increases, but the system is still stabilized before a full outage by a timely rollback or effective traffic shedding. Consequently, mitigative barriers include \emph{automatic rollback}, \emph{traffic shedding}, \emph{queue protection}, \emph{regional isolation}, and, if applicable, \emph{manual failover}. Already at this point it becomes apparent why the example is particularly suitable for Part~3: it contains not only a linear error chain, but several alternative damage trajectories, context-dependent barriers, and genuine intervention decisions.

\subsection{Instant-Payments Risk: From Bowtie to DAG}
Operational modeling in the Realtime Risk Studio does not begin with abstract categories, but with the question of which runtime states are later to be observed or influenced. A DAG is therefore derived from the Bowtie in which at least the following classes of nodes occur explicitly.

First, causes and context conditions are modeled as their own nodes, such as \emph{Faulty Change}, \emph{Routing Misconfiguration}, \emph{Validation Gate}, \emph{Peak Load Window}, and \emph{Observability Degraded}. Second, operational intermediate states are modeled that often remain implicit in the Bowtie, but are central for monitoring, for example \emph{Queue Saturation}, \emph{High Latency}, and \emph{Retry Storm}. Third, barriers are represented as Activation Nodes, such as \emph{Canary Rollout}, \emph{Automatic Rollback}, \emph{Traffic Shedding}, and \emph{Regional Isolation}. Fourth, a consequence node with alternative end states is modeled, for example \{\emph{safe}, \emph{degraded service}, \emph{partial outage}, \emph{transaction loss}\}. The state \emph{safe} represents the normal case of the system and is thus exactly the category that is missing in the Bowtie, but dominant in monitoring.

A concrete edge structure well suited to the example can be read as follows. \emph{Faulty Change} and \emph{Routing Misconfiguration}, together with \emph{Peak Load Window}, influence the probability of \emph{Service Degradation}. \emph{Validation Gate} and \emph{Canary Rollout} act inhibitingly on this transition. If \emph{Service Degradation} occurs, it influences \emph{Queue Saturation}; this in turn affects \emph{High Latency}; high latency increases the probability of a \emph{Retry Storm}. \emph{Traffic Shedding} and \emph{Queue Protection} intervene in this part of the model. \emph{Regional Isolation} and \emph{Automatic Rollback} act further to the right on the severity of the consequence, i.e., on whether the situation remains at \emph{degraded service}, tips into a \emph{partial outage}, or ends in \emph{transaction loss}. In this way, the Bowtie is transformed not into a merely differently drawn graph, but into a runtime model with explicit hypotheses about amplifiers, dampeners, and end states.

The choice of state spaces is also important for the example. Binary nodes are not sufficient everywhere. \emph{Peak Load Window} can sensibly be modeled as \{\emph{false}, \emph{true}\}, but \emph{Queue Saturation} is more realistically represented with graded states such as \{\emph{normal}, \emph{elevated}, \emph{critical}\}. The consequence node can likewise be multivalued. This multivaluedness is precisely what increases operational expressiveness: monitoring then answers not only \enquote{outage yes or no}, but probabilistically distinguishes between a still manageable degradation state and a business-critical damage state.

\begin{figure*}[t]
\centering
\includegraphics[width=0.97\textwidth]{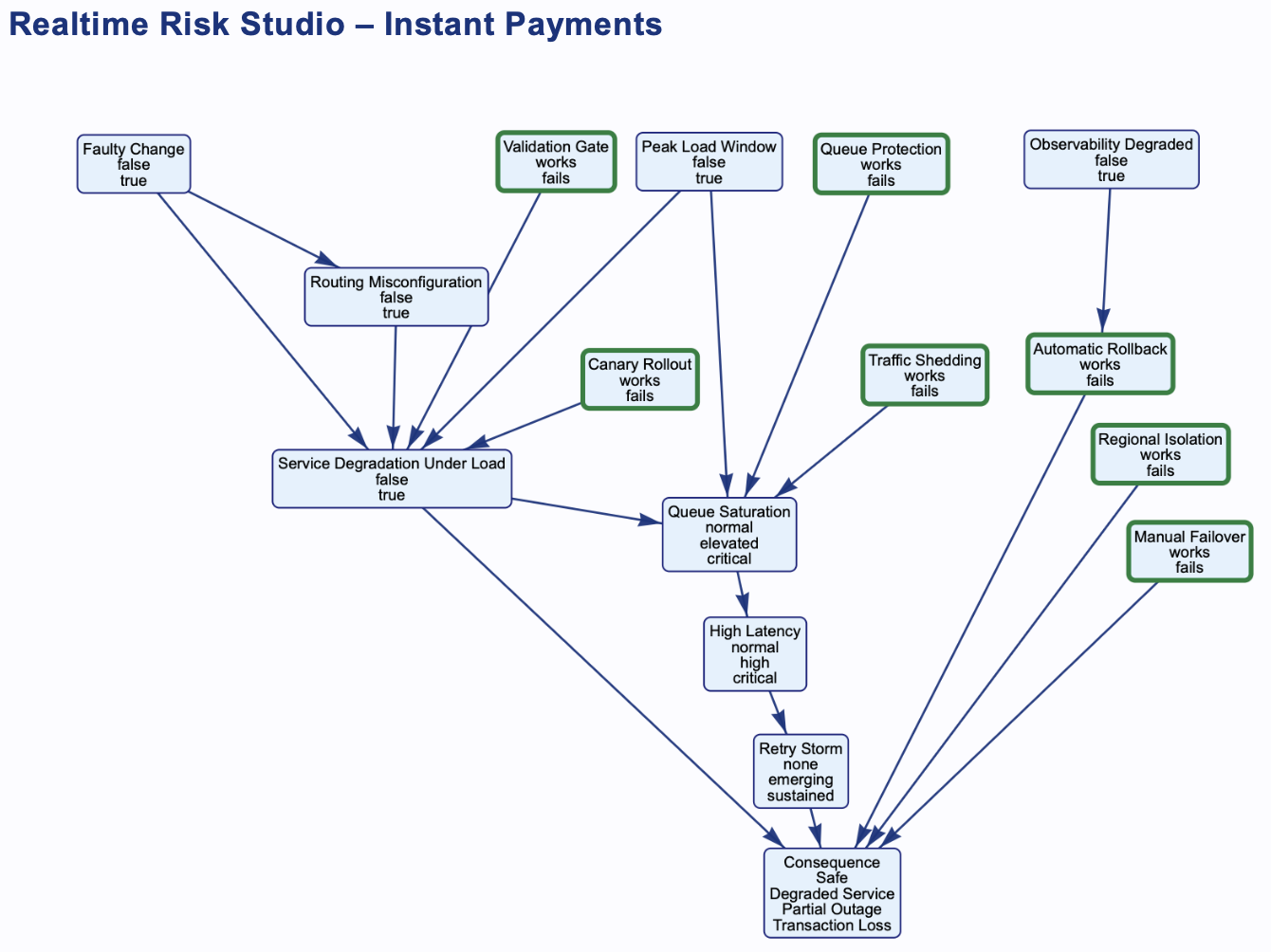}
\caption{Bowtie-derived runtime DAG of the instant-payments risk in the Realtime Risk Studio.}
\label{fig:studio-dag}
\end{figure*}

Figure~\ref{fig:studio-dag} shows the enriched Bowtie-derived DAG in the Realtime Risk Studio, including the explicit \emph{safe} state and several designated Activation Nodes (green border) that remain available for later intervention analysis.

\subsection{Building a Model in Realtime Risk Studio}
In the reference implementation, the case study would typically be modeled in four steps. First, the preventive structure to the left of the top event is taken over from the Bowtie: change, configuration, and load nodes are connected to \emph{Service Degradation}; barriers such as \emph{Validation Gate} and \emph{Canary Rollout} are created as Activation Nodes. Second, the right-hand part is added, i.e., the operational degradation chain with \emph{Queue Saturation}, \emph{High Latency}, \emph{Retry Storm}, and the mitigative barriers. Third, a common consequence node with \emph{safe} as the normal state is created. Fourth, integration information is prepared per node, such as REST sources for telemetry or REST notification endpoints for later push-based events.

In the instant-payments scenario in particular, this fourth stage is decisive. A node such as \emph{Peak Load Window} can be triggered by an external context indicator, a load classification, or another model. \emph{Queue Saturation} can be connected to a metrics or observability feed, \emph{Regional Isolation} to the status of regional traffic steering, and \emph{Automatic Rollback} to the actual state of the deployment automation. The model is thereby prepared from the outset so that it can later be fed not merely by manual assessment, but by runtime signals.

\begin{figure*}[t]
\centering
\includegraphics[width=0.97\textwidth]{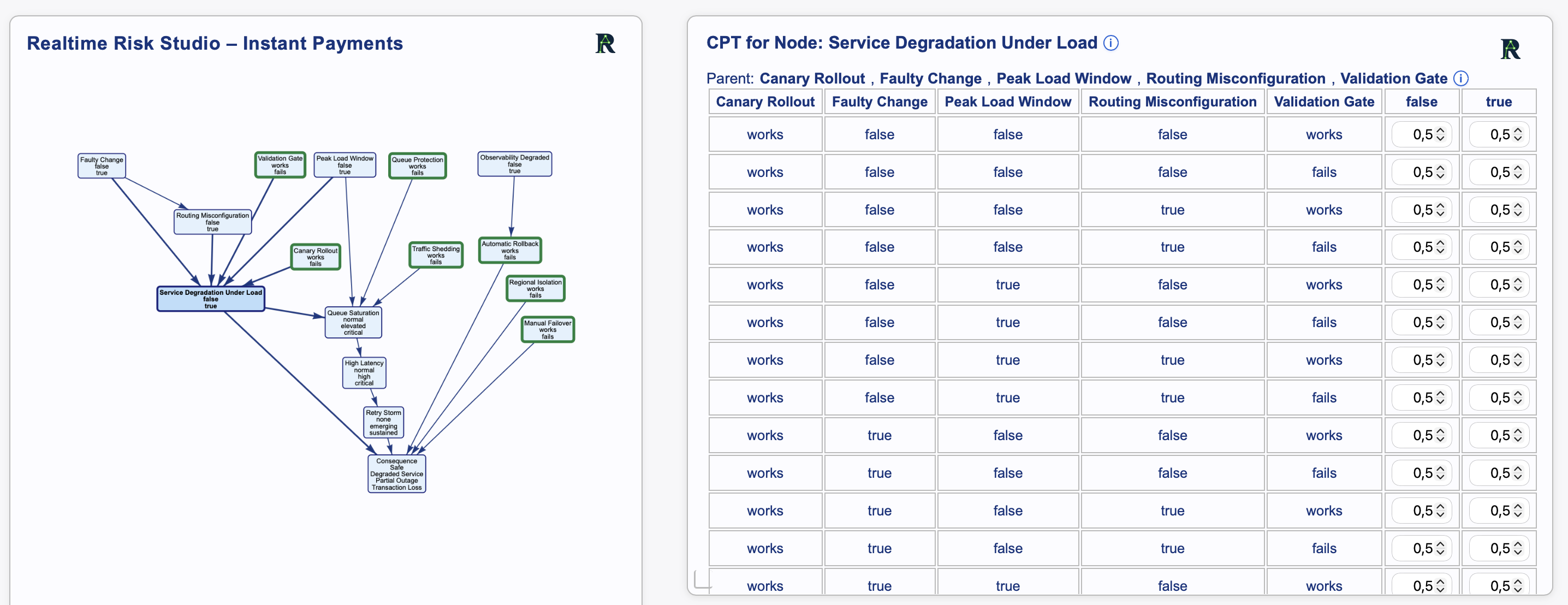}
\caption{DAG view and CPT editor for the node \emph{Service Degradation Under Load}.}
\label{fig:cpt-editor}
\end{figure*}

Once the graph structure is in place, individual CPTs can be inspected and refined directly in the studio. Figure~\ref{fig:cpt-editor} shows the DAG together with the CPT editor for \emph{Service Degradation Under Load}. Methodologically, this matters because it keeps the link between graphical structure and concrete conditional assumptions transparent: analysts can review parent configurations row by row, inspect how structure-derived questions map back into CPT cells, and still make targeted manual corrections where domain knowledge requires them.

\subsection{Probability Capture for the Case Study}
After structural modeling, the Probability Capture tool automatically generates the associated questions. In the example, it becomes particularly clear that different nodes require very different kinds of expert knowledge. For a node such as \emph{Service Degradation}, conditional questions arise, for example, of the form: \enquote{What is the probability that \texttt{Service Degradation=true}, given that \texttt{Faulty Change=true}, \texttt{Peak Load Window=true}, and \texttt{Canary Rollout=fails}?} For \emph{Queue Saturation}, questions are generated that jointly condition on load and degradation, and for the consequence node questions of the form: \enquote{What is the probability of \texttt{Transaction Loss}, given \texttt{Retry Storm=sustained}, \texttt{Regional Isolation=fails}, and \texttt{Automatic Rollback=fails}?}

Here it becomes clear why subgraph-based questionnaires are important. The platform and deployment perspective is especially relevant for \emph{Faulty Change}, \emph{Validation Gate}, \emph{Canary Rollout}, and \emph{Automatic Rollback}. System reliability engineering or operations roles can assess \emph{Queue Saturation}, \emph{High Latency}, \emph{Retry Storm}, \emph{Traffic Shedding}, and \emph{Regional Isolation} more robustly. Domain roles from payments or operations risk are more suitable for classifying the conditions of \emph{partial outage} and \emph{transaction loss} in domain terms. The tool can generate tokenized capture links precisely for this purpose, each containing only the relevant subgraph and thus only the domain-appropriate questions.

In the example, working with priors is also meaningful. For \emph{Peak Load Window}, a relatively empirical prior can be used because load windows in productive payment services are usually well known organizationally and technically. For \emph{Transaction Loss} or \emph{Complete Unavailability}, by contrast, a cautious prior with a smaller baseline value is plausible; such a prior neither claims that these states are common nor allows a single opinion to dominate the CPT immediately. In practice, this matters in the case study because precisely the rare, highly critical states often have the least historical data, yet the highest relevance for decision management.

\subsection{Noise Analysis of the Case Study}
The instant-payments gateway is particularly suitable for understanding the noise analysis. To make this visible without claiming an empirical benchmark, consider one illustrative elicitation round. For the comparatively well-anchored question \enquote{What is the probability that \texttt{Automatic Rollback=works} holds if \texttt{Faulty Change=true} and \texttt{Peak Load Window=true}?}, assume four deployment experts provide the values 0.78, 0.81, 0.79, and 0.84. In such a case, the dispersion is low, the arithmetic mean is 0.805, and a prior-regularized estimate remains close to that value. The important point is not the specific number, but the pattern: a question with shared operational meaning produces tight answers and therefore a comparatively stable CPT entry.

The situation is different for questions with hidden context heterogeneity. For the question \enquote{What is the probability that \texttt{Retry Storm=sustained} occurs if \texttt{High Latency=true}?}, consider instead the illustrative answers 0.20, 0.35, 0.62, and 0.78. Here the wide spread is not merely a psychological artifact, but a signal that the conditioning context has not yet been anchored sufficiently in domain terms. Some experts may implicitly think of disciplined client software that backs off automatically, others of heterogeneous systems without effective retry limitation. The tool turns this divergence into a modeling signal: either the question is sharpened, for example by restricting it to a specific channel, a specific load window, and a specific client type, or the model is extended by an additional node such as \emph{Client Retry Discipline}. In the reference implementation, questions of this kind are precisely the cases in which an analyst may lower the project-level Expert $\kappa$ for a more cautious Expert-Consensus aggregation, while an optional half-life can be activated when recency effects are relevant. In this way, the noise analysis serves not only response statistics, but model improvement itself.

\begin{figure*}[t]
\centering
\includegraphics[width=0.97\textwidth]{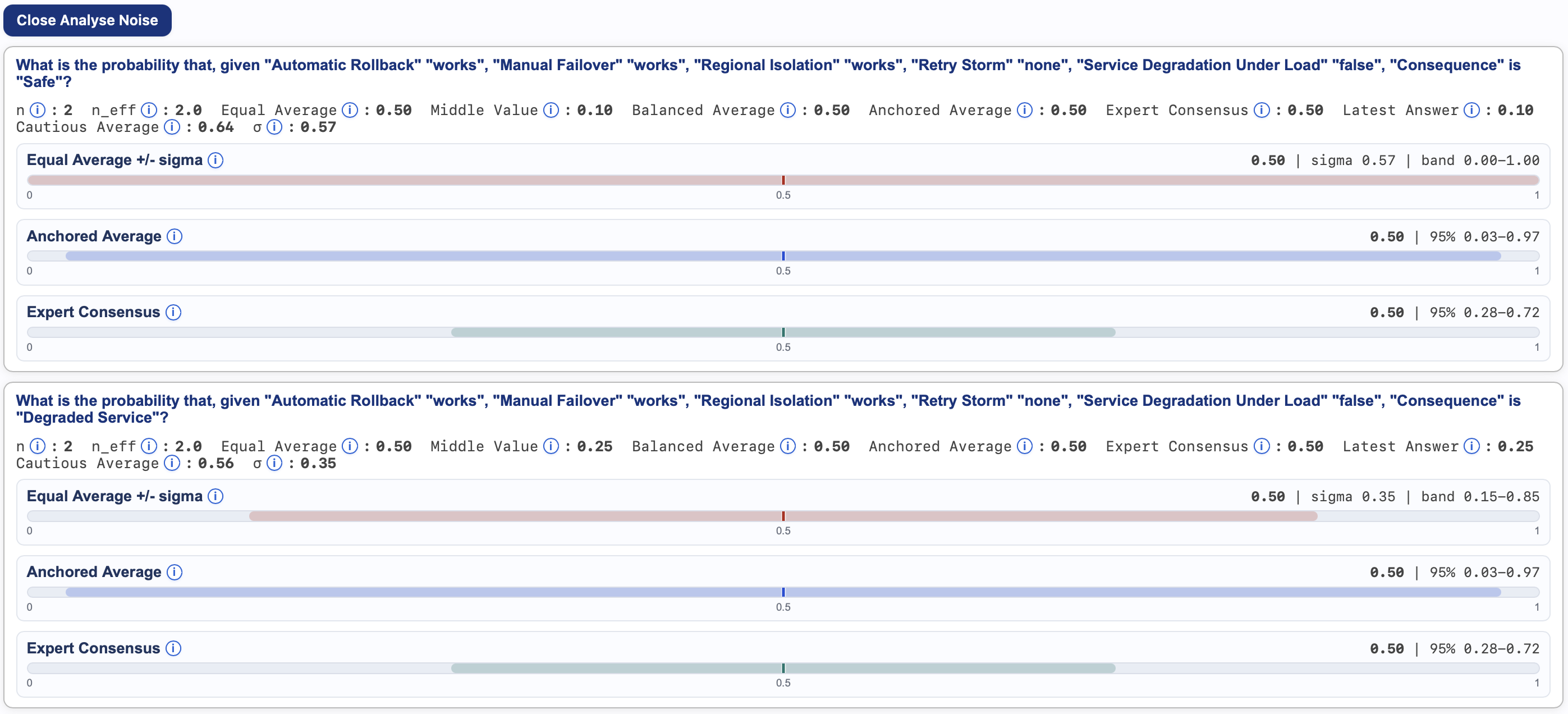}
\caption{Noise-analysis view for conditional probability questions in the instant-payments example.}
\label{fig:noise-analysis}
\end{figure*}

Figure~\ref{fig:noise-analysis} illustrates this for the instant-payments example. The screenshot makes visible how the same conditional question is summarized by several estimators and how the spread-sensitive views expose whether a CPT entry is comparatively stable or still weakly anchored in shared expert judgment.

\subsection{Runtime Use: Three Operational Situation Cuts}

For demonstrating the added value during runtime, without claiming a statistical validation, Table~\ref{tab:situation-cuts} reports the posterior probabilities of the consequence node at three operational situation cuts in the case study. The values are not measured incident frequencies; they simply instantiate the BN in a way that makes the monitoring logic explicit.

\begin{table}[t]
\caption{Illustrative consequence-state posteriors for three operational situation cuts in the instant-payments example.}
\label{tab:situation-cuts}
\centering
\scriptsize
\setlength{\tabcolsep}{3pt}
\begin{tabularx}{\columnwidth}{@{}l>{\raggedright\arraybackslash}Xrrrr@{}}
\toprule
Cut & Evidence & Safe & Degr. & Part. & Loss \\
\midrule
1 & Peak load, normal queue, low latency, rollback available & 0.94 & 0.05 & 0.01 & 0.00 \\
2 & Faulty change, elevated queue, high latency, retry state unclear & 0.46 & 0.38 & 0.13 & 0.03 \\
3 & Critical queue, sustained retry storm, isolation fails, rollback fails & 0.06 & 0.22 & 0.45 & 0.27 \\
\bottomrule
\end{tabularx}
\end{table}

Cut~1 corresponds to tense but still controlled operation. Load is present, yet the observable degradation indicators remain normal. In the illustrative parametrization, the posterior of the consequence node is therefore still dominated by \emph{safe} with probability 0.94. This is exactly what matters for a realtime model: load alone must not automatically be interpreted as a crisis.

Cut~2 represents the transition into a genuinely unstable situation. Once \emph{Faulty Change=true} and rising latency-related evidence are introduced, the posterior mass shifts away from \emph{safe} and toward \emph{degraded service}. The important point is not any single number, but that the model returns a state distribution rather than a binary alarm. Operations therefore sees that the system has not yet fully tipped into loss, but has already moved into a materially different risk state.

Cut~3 represents a clearly escalated incident picture. Under critical queue saturation, a sustained retry storm, and failing mitigation, the illustrative posterior of \emph{Transaction Loss} rises to 0.27, while \emph{partial outage} becomes the most likely state at 0.45. At this point, monitoring turns into decision support: the model no longer merely describes that the situation is bad, but provides a quantified baseline against which interventions can be compared.

\subsection{Causal Analysis and Intervention Search}
Once a model has been parametrized, a causal analysis can be performed. In the case study, several questions are particularly natural. One first question is: what causal effect does \emph{Faulty Change} have on \emph{Transaction Loss}? For this purpose, the reference implementation can determine minimal or all backdoor adjustment sets and thereby make visible which variables should be controlled in a sound effect analysis, for example \emph{Peak Load Window} or other context variables that influence both the occurrence of degradation and the severity of the consequence.

Particularly relevant for operations is the do-analysis on Activation Nodes. Using the illustrative Cut~3 evidence from Table~\ref{tab:situation-cuts} as the baseline, the posterior target probability is $P(\texttt{Transaction Loss=true})=0.27$. For the same instantiated model, an intervention search can then compare candidate actions under identical evidence. In one ranking, setting \texttt{do(Automatic Rollback=works)} reduces the target probability to 0.18, \texttt{do(Traffic Shedding=works)} reduces it to 0.12, and \texttt{do(Regional Isolation=works)} reduces it to 0.09. In this illustrative configuration, \emph{Regional Isolation} therefore ranks ahead of \emph{Traffic Shedding}, which in turn ranks ahead of \emph{Automatic Rollback}.

This ranking is not presented as a universal rule for instant‑payment incidents; it depends on the evidence and the parametrization. Under a different evidence pattern, for example, earlier in the escalation chain and with stronger evidence for a faulty rollout as the primary driver, \emph{Automatic Rollback} could become the highest-impact intervention. This context dependence is the advantage of a causal model: it replaces intuitions with an evidence-conditioned comparison of candidate measures.

\begin{figure*}[t]
\centering
\includegraphics[width=0.97\textwidth]{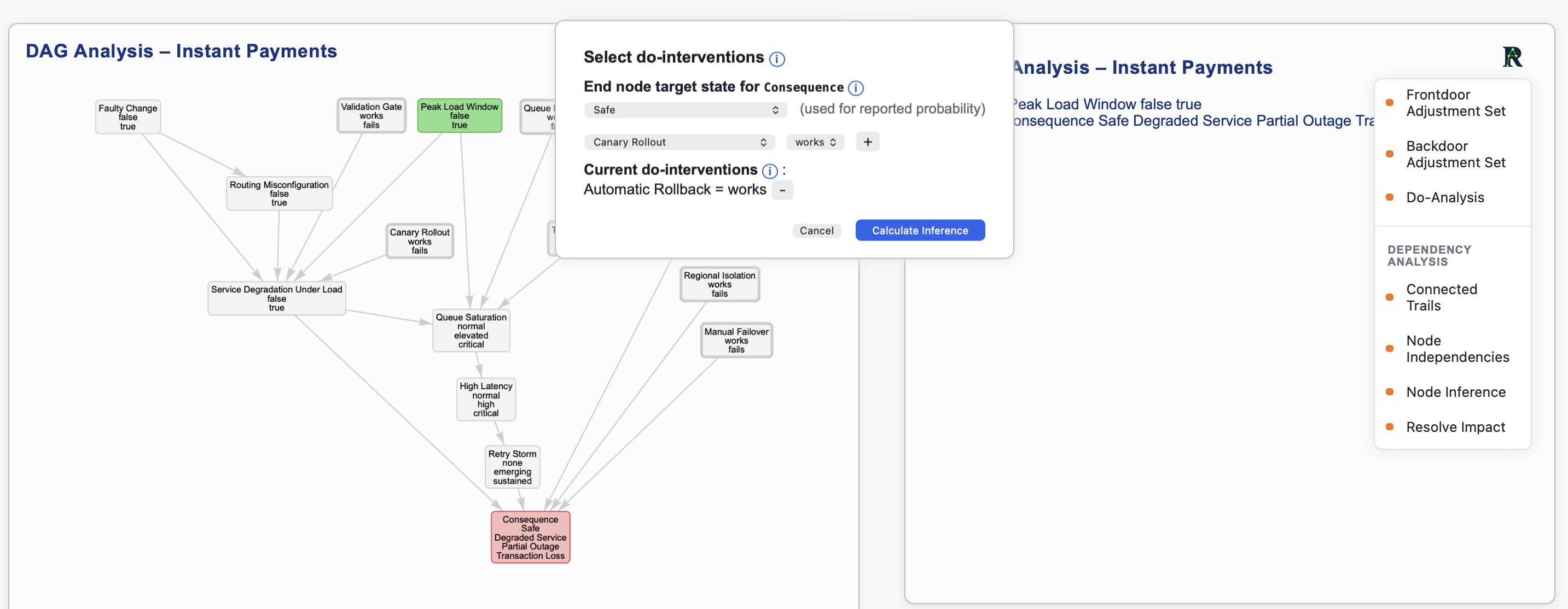}
\caption{Causal-analysis view for the instant-payments case.}
\label{fig:causal-analysis}
\end{figure*}

Figure~\ref{fig:causal-analysis} shows the corresponding causal-analysis view. In the screenshot, a target end state for the consequence node is selected, candidate do-interventions can be instantiated, and the resulting analysis panels expose adjustment-set information as well as intervention-oriented reasoning paths side by side.

\subsection{Outlook: Monitoring and Decision Hub}
The case study thus demonstrates how the proposed methods and reference implementation are intended to work operationally. The Realtime Risk Studio structures the risk not only semantically, but in the form of observable nodes and actionable protective functions. The Probability Capture tool translates this structure into questionnaires for experts and makes dispersion as well as prior effects explicitly visible. The causal analysis finally provides a formal basis for deciding among measures under given evidence.

For our Monitoring and Decision Hub, planned for Part~4 in this publication series, this is directly applicable in the instant-payments gateway. Telemetry on queue lengths, latencies, error rates, rollback status, and regional state can flow into the model as evidence. The model keeps the \emph{safe} state explicitly present, thus distinguishing normal operation from genuine degradation, and can at the same time prioritize, under crisis evidence, those Activation Nodes that have the greatest expected effect on the target variable \emph{transaction loss}. This risk cannot be completely eliminated by design, which makes it a useful illustration of the practical value of the proposed method.

\section{Discussion}
The presented methods and toolchain show that realtime risk management neither begins nor ends with a single method. Heatmaps, Bowties, DAGs, Probability Capture, and causal analysis address different levels of tasks. The added value of the Hagenberg Risk Management Process therefore lies not in replacing one of the established methods, but in connecting them in an operationally consistent sequence. The contribution of this paper is the integrated bridge from a Bowtie model to a runtime-capable DAG with explicit safe-state semantics, estimator-aware CPT elicitation, and intervention-oriented analysis on the same model.

The most important methodological decision of this paper is the explicit addition of the \emph{safe state}. From a Bayesian-network perspective, this addition may appear almost self-evident, but in practical risk management it has far-reaching consequences. Only through this step is the dominant operating case represented formally. Without the \emph{safe} state, the model would distinguish only among manifestations of damage in the consequence logic, but not between \emph{normal and harmful}. For monitoring, this distinction is central.

A second strength is the clean treatment of barriers. In classic Bowties, barriers often remain a semantically strong, but formally underdetermined category. In the approach described here, they are modeled as state nodes and additionally marked as Activation Nodes. This makes them observable, parameterizable, and treatable as intervention points in causal analyses. This dual role is particularly valuable for the transition from risk argumentation to decision management.

A third strength lies in the noise analysis implemented in the Probability Capture tool. Maintaining conditional probability tables (CPTs) in a Bayesian network is a difficult task, and the tool tackles this problem directly: questions are generated from model structure, tailored to subgraphs, complemented with priors, and evaluated with respect to dispersion. Parametrization thereby becomes an independent quality process rather than a hidden appendix to the graph design.

At the same time, limitations must be openly discussed. First, the quality of a model continues to depend strongly on the quality of its structural assumptions. A cleanly visualized or mathematically correct DAG does not compensate for conceptually wrong edges. Second, expert estimates remain subjective despite dispersion analysis. Prior-regularized aggregation can stabilize sparse inputs, but it does not replace systematic empirical calibration. Third, the estimator family used here serves different operational purposes and should not be interpreted as equally neutral statistical estimators. In particular, Cautious Average is best understood as a conservative policy heuristic, the Anchored Average refers to a prior-regularized mean, and the Expert Consensus is a bounded-support latent-consensus model with a concentration parameter for updating priors with expert responses. Fourth, the current DAG does not represent true temporal feedback loops with cycles; for strongly time-dependent control loops, an extension toward dynamic Bayesian networks may be useful~\cite{khakzad2012,khakzad2013}. Fifth, causal analysis is only as robust as the substantive plausibility of the modeled causal structure and the adequacy of the available evidence.

A further limitation concerns the evidential status of the case study. The instant-payments example is deliberately domain-realistic and useful for illustrating the workflow, but it does not yet provide a quantitative benchmark, a calibration study, or a head-to-head comparison with alternative elicitation and runtime-monitoring strategies. Such evaluation would be a next step. 

Nevertheless, the combination is particularly well suited to the target class of material, observation-relevant risks. Precisely where risks can neither be trivially accepted nor fully designed away, a formal model is needed that connects structure, uncertainty, evidence, and action-oriented reasoning. It is exactly this gap that the Realtime Risk Studio addresses.

\section{Conclusion}
The present paper continues the Hagenberg Risk Management Process in its third step toward operationally usable realtime risk management. The initial observation is that a Bowtie is a strong instrument for the semantic structuring of risks, but not a sufficient modeling basis for runtime monitoring. The reason lies not only in its often static and qualitative use, but above all in the fact that the normal state of the system is not modeled explicitly. Yet this very state must be represented if monitoring is to distinguish normal operation from genuine degradation. The core contribution of this paper is therefore not a single new estimator or causal criterion, but a workflow that connects Bowtie semantics, safe-state operationalization, estimator-aware CPT elicitation, and intervention-oriented reasoning.

With the Realtime Risk Studio as our reference implementation, the Bowtie-to-DAG mapping including a \emph{safe} state, the Probability Capture tool with dispersion analysis and prior-regularized aggregation, and the causal-analysis mode, an end-to-end toolchain is presented that addresses this gap. The system combines semantic risk modeling with probabilistic state description, evidence-oriented runtime use, and intervention analysis. The instant-payments gateway case study shows how a material, DORA-related risk can be transformed into an observable and intervention-ready model.

Overall, the paper argues that the proposed combination is particularly suitable for risks with context-dependent behavior, limited acceptability, persistent uncertainty, and the need for timely operational reaction. Further empirical work is needed in order to benchmark the elicitation procedures, calibrate the estimators, and validate operational performance under real incident or near-miss data. Even with this limitation, the combination of Bowtie semantics, DAG operationalization, Probability Capture, and causal analysis provides a coherent methodological foundation for realtime risk management.

\paragraph*{Artifact Availability} The current Realtime Risk Studio is part of Riskomat, the reference implementation of the Hagenberg Risk Management Process. For more information about the implemented tools and the integrated risk management platform, feel free to email us at \url{info@kompilomat.com}.

\section*{AI Disclosure}

As our scientific work focuses on the integration of AI-based methods into various types of processes, particularly for security-related applications, large language models were used to support the revision of this manuscript. Their use was limited to editorial assistance, including the refinement of figures as well as improvements to clarity, wording, and error correction in the main text. The authors take full responsibility for the content of this work.


\begin{thebibliography}{99}

\bibitem{hermann2026part1}
E.~Hermann and H.~Lampesberger, ``Hagenberg Risk Management Process (Part~1): Multidimensional Polar Heatmaps for Context-Sensitive Risk Analysis,'' \emph{arXiv:2601.07644}, 2026.

\bibitem{hermann2026part2}
E.~Hermann and H.~Lampesberger, ``Hagenberg Risk Management Process (Part~2): From Context-Sensitive Triage to Case Analysis With Bowtie and Bayesian Networks,'' \emph{arXiv:2602.19270}, 2026.

\bibitem{dora2022}
European Union, ``Regulation (EU) 2022/2554 of the European Parliament and of the Council of 14 December 2022 on digital operational resilience for the financial sector (DORA),'' \emph{Official Journal L 333}, 2022.

\bibitem{nist800137}
K.~L.~Dempsey, N.~S.~Chawla, L.~A.~Johnson, R.~Johnston, A.~C.~Jones, A.~D.~Orebaugh, M.~A.~Scholl, and K.~M.~Stine, ``Information Security Continuous Monitoring (ISCM) for Federal Information Systems and Organizations,'' NIST Special Publication 800-137, 2011.

\bibitem{deruijter2016}
A.~de~Ruijter and F.~Guldenmund, ``The bowtie method: A review,'' \emph{Safety Science}, vol.~88, pp.~211--218, 2016.

\bibitem{iec31010}
International Electrotechnical Commission, ``IEC 31010:2019 Risk management---Risk assessment techniques,'' IEC Standard, 2019.

\bibitem{khakzad2013}
N.~Khakzad, F.~Khan, and P.~Amyotte, ``Dynamic safety analysis of process systems by mapping bow-tie into Bayesian network,'' \emph{Process Safety and Environmental Protection}, vol.~91, no.~1--2, pp.~46--53, 2013.

\bibitem{khakzad2012}
N.~Khakzad, F.~Khan, and P.~Amyotte, ``Dynamic risk analysis using bow-tie approach,'' \emph{Reliability Engineering \& System Safety}, vol.~104, pp.~36--44, 2012.

\bibitem{zurheide2021}
F.~T.~Zurheide, E.~Hermann, and H.~Lampesberger, ``pyBNBowtie: Python library for Bow-Tie Analysis based on Bayesian Networks,'' \emph{Procedia Computer Science}, vol.~180, pp.~344--351, 2021.

\bibitem{koller2009}
D.~Koller and N.~Friedman, \emph{Probabilistic Graphical Models: Principles and Techniques}. Cambridge, MA, USA: MIT Press, 2009.

\bibitem{fenton2012}
N.~Fenton and M.~Neil, \emph{Risk Assessment and Decision Analysis with Bayesian Networks}, 2nd~ed. Boca Raton, FL, USA: CRC Press, 2018.

\bibitem{pearl2009}
J.~Pearl, \emph{Causality: Models, Reasoning, and Inference}, 2nd~ed. Cambridge, U.K.: Cambridge Univ. Press, 2009.

\bibitem{textor2016}
J.~Textor, B.~van~der~Zander, M.~S.~Gilthorpe, M.~Li\'skiewicz, and G.~T.~H.~Ellison, ``Robust causal inference using directed acyclic graphs: the R package `dagitty','' \emph{International Journal of Epidemiology}, vol.~45, no.~6, pp.~1887--1894, 2016.

\bibitem{ankan2024}
A.~Ankan and J.~Textor, ``pgmpy: A Python Toolkit for Bayesian Networks,'' \emph{Journal of Machine Learning Research}, vol.~25, no.~265, pp.~1--8, 2024.

\bibitem{scutari2010}
M.~Scutari, ``Learning Bayesian Networks with the bnlearn R Package,'' \emph{Journal of Statistical Software}, vol.~35, no.~3, pp.~1--22, 2010.

\bibitem{druzdzel1999}
M.~J.~Druzdzel, ``SMILE: Structural Modeling, Inference, and Learning Engine and GeNIe: A Development Environment for Graphical Decision-Theoretic Models,'' in \emph{Proc. AAAI/IAAI}, 1999, pp.~902--903.

\bibitem{madsen2003}
A.~L.~Madsen, M.~Lang, U.~B.~Kj\ae rulff, and F.~Jensen, ``The Hugin Tool for Learning Bayesian Networks,'' in \emph{Symbolic and Quantitative Approaches to Reasoning with Uncertainty}, T.~D.~Nielsen and N.~L.~Zhang, Eds. Berlin, Germany: Springer, 2003, pp.~594--605.

\bibitem{iso31000}
International Organization for Standardization, ``ISO 31000:2018 Risk management---Guidelines,'' ISO Standard, 2018.

\bibitem{badreddine2013}
A.~Badreddine and N.~Ben~Amor, ``A Bayesian approach to construct bow tie diagrams for risk evaluation,'' \emph{Process Safety and Environmental Protection}, vol.~91, no.~3, pp.~159--171, 2013.

\bibitem{vairo2022}
T.~Vairo, A.~C.~Benvenuto, A.~Tedeschi, A.~P.~Reverberi, and B.~Fabiano, ``Make Bow-tie Dynamic by Rethinking It as a Hierarchical Bayesian Network: Dynamic Risk Assessment of an LNG Bunkering Operation,'' \emph{Chemical Engineering Transactions}, vol.~91, pp.~277--282, 2022.

\bibitem{naderpour2014}
M.~Naderpour, J.~Lu, and G.~Zhang, ``A situation risk awareness approach for process systems safety,'' \emph{Safety Science}, vol.~64, pp.~173--189, 2014.

\bibitem{xing2019}
J.~Xing, Z.~Zeng, and E.~Zio, ``A framework for dynamic risk assessment with condition monitoring data and inspection data,'' \emph{Reliability Engineering \& System Safety}, vol.~191, p.~106552, 2019.

\bibitem{moradi2022}
R.~Moradi, S.~Cofre-Martel, E.~Lopez~Droguett, M.~Modarres, and K.~M.~Groth, ``Integration of deep learning and Bayesian networks for condition and operation risk monitoring of complex engineering systems,'' \emph{Reliability Engineering \& System Safety}, vol.~222, p.~108433, 2022.

\bibitem{moradi2023}
R.~Moradi, A.~Ruiz-Tagle~Palazuelos, E.~Lopez~Droguett, and K.~M.~Groth, ``Toward a framework for risk monitoring of complex engineering systems with online operational data: A deep learning-based solution,'' \emph{Proceedings of the Institution of Mechanical Engineers, Part O: Journal of Risk and Reliability}, vol.~237, no.~5, pp.~910--921, 2023.

\bibitem{zio2018}
E.~Zio, ``The future of risk assessment,'' \emph{Reliability Engineering \& System Safety}, vol.~177, pp.~176--190, 2018.

\bibitem{weber2012}
P.~Weber, G.~Medina-Oliva, C.~Simon, and B.~Iung, ``Overview on Bayesian networks applications for dependability, risk analysis and maintenance areas,'' \emph{Engineering Applications of Artificial Intelligence}, vol.~25, no.~4, pp.~671--682, 2012.

\bibitem{langseth2007}
H.~Langseth and L.~Portinale, ``Bayesian networks in reliability,'' \emph{Reliability Engineering \& System Safety}, vol.~92, no.~1, pp.~92--108, 2007.

\bibitem{kabir2019}
S.~Kabir and Y.~Papadopoulos, ``Applications of Bayesian networks and Petri nets in safety, reliability, and risk assessments: A review,'' \emph{Safety Science}, vol.~115, pp.~154--175, 2019.

\bibitem{george2021}
P.~G.~George and V.~R.~Renjith, ``Evolution of Safety and Security Risk Assessment methodologies towards the use of Bayesian Networks in Process Industries,'' \emph{Process Safety and Environmental Protection}, vol.~149, pp.~758--775, 2021.

\bibitem{cai2013}
B.~Cai, Y.~Liu, Z.~Liu, X.~Tian, Y.~Zhang, and R.~Ji, ``Application of Bayesian Networks in Quantitative Risk Assessment of Subsea Blowout Preventer Operations,'' \emph{Risk Analysis}, vol.~33, no.~7, pp.~1293--1311, 2013.

\bibitem{aquaro2010}
V.~Aquaro, M.~Bardoscia, R.~Bellotti, A.~Consiglio, F.~De~Carlo, and G.~Ferri, ``A Bayesian Networks approach to Operational Risk,'' \emph{Physica A: Statistical Mechanics and its Applications}, vol.~389, no.~8, pp.~1721--1728, 2010.

\bibitem{cornwell2023}
N.~Cornwell, C.~Bilson, A.~Gepp, S.~Stern, and B.~J.~Vanstone, ``Modernising operational risk management in financial institutions via data-driven causal factors analysis: A pre-registered study,'' \emph{Pacific-Basin Finance Journal}, vol.~79, p.~102011, 2023.

\bibitem{ohagan2006}
A.~O'Hagan, C.~E.~Buck, A.~Daneshkhah, J.~R.~Eiser, P.~H.~Garthwaite, D.~J.~Jenkinson, J.~E.~Oakley, and T.~Rakow, \emph{Uncertain Judgements: Eliciting Experts' Probabilities}. Chichester, U.K.: Wiley, 2006.

\bibitem{werner2017}
C.~Werner, T.~Bedford, R.~M.~Cooke, A.~M.~Hanea, and O.~Morales-N\'apoles, ``Expert judgement for dependence in probabilistic modelling: A systematic literature review and future research directions,'' \emph{European Journal of Operational Research}, vol.~258, no.~3, pp.~801--819, 2017.

\bibitem{pitchforth2013}
J.~Pitchforth and K.~Mengersen, ``A proposed validation framework for expert elicited Bayesian Networks,'' \emph{Expert Systems with Applications}, vol.~40, no.~1, pp.~162--167, 2013.

\bibitem{leonelli2023}
M.~Leonelli, R.~Ramanathan, and R.~L.~Wilkerson, ``Sensitivity and robustness analysis in Bayesian networks with the bnmonitor R package,'' \emph{Knowledge-Based Systems}, vol.~278, p.~110882, 2023.

\bibitem{pearl1988}
J.~Pearl, \emph{Probabilistic Reasoning in Intelligent Systems: Networks of Plausible Inference}. San Mateo, CA, USA: Morgan Kaufmann, 1988.

\bibitem{darwiche2009}
A.~Darwiche, \emph{Modeling and Reasoning with Bayesian Networks}. Cambridge, U.K.: Cambridge Univ. Press, 2009.

\bibitem{lauritzen1988}
S.~L.~Lauritzen and D.~J.~Spiegelhalter, ``Local computations with probabilities on graphical structures and their application to expert systems,'' \emph{Journal of the Royal Statistical Society: Series B}, vol.~50, no.~2, pp.~157--224, 1988.

\bibitem{spirtes2000}
P.~Spirtes, C.~Glymour, and R.~Scheines, \emph{Causation, Prediction, and Search}, 2nd~ed. Cambridge, MA, USA: MIT Press, 2000.

\bibitem{spirtes2016}
P.~Spirtes and K.~Zhang, ``Causal discovery and inference: concepts and recent methodological advances,'' \emph{Applied Informatics}, vol.~3, Art.~no.~3, 2016.

\bibitem{glymour2019}
C.~Glymour, K.~Zhang, and P.~Spirtes, ``Review of causal discovery methods based on graphical models,'' \emph{Frontiers in Genetics}, vol.~10, p.~524, 2019.

\bibitem{kalisch2012}
M.~Kalisch, M.~M\"achler, D.~Colombo, M.~H.~Maathuis, and P.~B\"uhlmann, ``Causal Inference Using Graphical Models with the R Package pcalg,'' \emph{Journal of Statistical Software}, vol.~47, no.~11, pp.~1--26, 2012.

\bibitem{tikka2017}
S.~Tikka and J.~Karvanen, ``Identifying Causal Effects with the R Package causaleffect,'' \emph{Journal of Statistical Software}, vol.~76, no.~12, pp.~1--30, 2017.

\bibitem{zheng2024}
Y.~Zheng, B.~Huang, W.~Chen, J.~Ramsey, M.~Gong, R.~Cai, S.~Shimizu, P.~Spirtes, and K.~Zhang, ``Causal-learn: Causal Discovery in Python,'' \emph{Journal of Machine Learning Research}, vol.~25, no.~60, pp.~1--7, 2024.

\bibitem{kalainathan2020}
D.~Kalainathan, O.~Goudet, and R.~Dutta, ``Causal Discovery Toolbox: Uncovering causal relationships in Python,'' \emph{Journal of Machine Learning Research}, vol.~21, no.~37, pp.~1--5, 2020.

\bibitem{ikeuchi2023}
T.~Ikeuchi, M.~Ide, Y.~Zeng, T.~N.~Maeda, and S.~Shimizu, ``Python package for causal discovery based on LiNGAM,'' \emph{Journal of Machine Learning Research}, vol.~24, no.~14, pp.~1--8, 2023.

\bibitem{huntingtonklein2021}
N.~Huntington-Klein, \emph{The Effect: An Introduction to Research Design and Causality}. Boca Raton, FL, USA: CRC Press, 2021.

\bibitem{kahneman2021}
D.~Kahneman, O.~Sibony, and C.~R.~Sunstein, \emph{Noise: A Flaw in Human Judgment}. New York, NY, USA: Little, Brown Spark, 2021.

\bibitem{taleb2007}
N.~N.~Taleb, ``Black Swans and the Domains of Statistics,'' \emph{The American Statistician}, vol.~61, no.~3, pp.~198--200, 2007.

\bibitem{taleb2010}
N.~N.~Taleb, \emph{The Black Swan: The Impact of the Highly Improbable}, 2nd~ed. New York, NY, USA: Random House Trade Paperbacks, 2010.

\end{thebibliography}
\end{document}